%% file: ICL.tex
\documentclass[10pt,conference]{IEEEtran}

\IEEEoverridecommandlockouts
\usepackage{cite}
\usepackage{amsmath,amssymb,amsfonts}
\usepackage{graphicx}
\usepackage{textcomp}
\usepackage{xcolor}

\usepackage{algorithm}
\usepackage{algorithmic}
\usepackage{booktabs}
\usepackage{multirow}
\usepackage{bbding}

\usepackage{makecell}
\usepackage{bigstrut}

\usepackage{graphicx}
\usepackage{subfigure} 
\usepackage{hyperref }

\usepackage{multirow,multicol}
\usepackage[most]{tcolorbox}

\usepackage{utfsym}
\usepackage{soul}

\usepackage{makecell}
\usepackage{enumitem}
\usepackage[numbers]{natbib}
\usepackage{wasysym}

\def\BibTeX{{\rm B\kern-.05em{\sc i\kern-.025em b}\kern-.08em
    T\kern-.1667em\lower.7ex\hbox{E}\kern-.125emX}}

\begin{document}

\title{ZC$^3$: Zero-Shot Cross-Language Code Clone Detection
}

\author{
\IEEEauthorblockN{Jia Li}
\IEEEauthorblockA{{Key Lab of High Confidence Software} \\
Technology, MoE (Peking University)\\
Beijing, China \\
lijiaa@pku.edu.cn}

\\

\IEEEauthorblockN{ Fang Liu}
\IEEEauthorblockA{{Beihang University} \\
School of Computer Science \\
Beijing, China \\
fangliu@buaa.edu.cn }

\and

\IEEEauthorblockN{\textsuperscript{} Chongyang Tao}
\IEEEauthorblockA{{School of Computer Science} \\
Peking University \\
Beijing, China \\
chongyangtao@gmail.com}

\\

\IEEEauthorblockN{\textsuperscript{} Jia Li \male}
\IEEEauthorblockA{{Key Lab of High Confidence Software} \\
Technology, MoE (Peking University)\\
Beijing, China \\
lijia@stu.pku.edu.cn}

\and

\IEEEauthorblockN{\textsuperscript{} Zhi Jin* \thanks{ * Corresponding authors}}
\IEEEauthorblockA{{Key Lab of High Confidence Software} \\
Technology, MoE (Peking University)\\
Beijing, China \\
zhijin@pku.edu.cn}

\\ 

\IEEEauthorblockN{\textsuperscript{} Ge Li*}
\IEEEauthorblockA{{Key Lab of High Confidence Software} \\
Technology, MoE (Peking University) \\
Beijing, China \\
lige@pku.edu.cn}
}

\maketitle

\begin{abstract}



Developers introduce code clones to improve programming productivity. Many existing studies have achieved impressive performance in monolingual code clone detection.
However, during software development, more and more developers write semantically equivalent programs with different languages to support different platforms and help developers translate projects from one language to another. Considering that collecting cross-language parallel data, especially for low-resource languages, is expensive and time-consuming, how designing an effective cross-language model that does not rely on any parallel data is a significant problem.
In this paper, we propose a novel method named ZC$^3$ for \underline{Z}ero-shot \underline{C}ross-language \underline{C}ode \underline{C}lone detection. ZC$^3$ designs the contrastive snippet prediction to form an isomorphic representation space among different programming languages. Based on this, ZC$^3$ exploits domain-aware learning and cycle consistency learning to further constrain the model to generate representations that are aligned among different languages meanwhile are diacritical for different types of clones. To evaluate our approach, we conduct extensive experiments on four representative cross-language clone detection datasets. Experimental results show that ZC$^3$ outperforms the state-of-the-art baselines by 67.12\%, 51.39\%, 14.85\%, and 53.01\% on the MAP score, respectively. We further investigate the representational distribution of different languages and discuss the effectiveness of our method.

\end{abstract}


\input{1_introduction}

\input{2_related}

\input{3_method}
\input{4_experiments}

\input{7_conclusion}

\bibliographystyle{IEEEtranN}
\bibliography{IEEEabrv,ICL}

\end{document}

%% file: 1_introduction.tex
\section{Introduction}
Code clones are similar or selfsame source codes that have the same functionality. 
Given a source code as the query, code clone detection aims to retrieve codes with the same semantics from a collection of candidates. 
With the increasing number of programming language types and the rapid expansion of open-source codes, developers are facing a growing need for detecting code clones across different languages. 
Programmers need to develop semantically equivalent but linguistically distinct programs to support diverse platforms, such as Apps in C/C\# for a Windows system, Java for an Android system, and Objective-C for a Mac system~\cite{tao2022c4, cheng2016mining, nafi2019clcdsa}. 
Meanwhile, when requested to implement a specific function with an unfamiliar programming language, cross-language clone detection can help developers retrieve corresponding clones by using their proficient language programs and accomplish program migration. 
Besides, duplicated source codes increase the complexity of software maintenance and might introduce fault propagation if code clones change inconsistently. 
Therefore, the ability to automatically detect cross-language clones is significant for efficiently maintaining software systems and helps developers translate codes from one language to another by retrieving source codes with the same semantics.


\begin{figure*}[htbp]
	\centering
        \setlength{\abovecaptionskip}{0.1cm} 
	\vspace{-0.35cm} 
	\subfigtopskip=0pt 
	\subfigcapskip=0pt 
        \subfigure[PHP]{
	\begin{minipage}{0.33\linewidth}
		\centering
		\includegraphics[width=0.95\linewidth]{figure/php.pdf}
		\label{php} 
	\end{minipage}
        } \hspace{-5mm}
        \subfigure[Java]{
	\begin{minipage}{0.33\linewidth}
		\centering
		\includegraphics[width=0.95\linewidth]{figure/java.pdf}
		\label{java}
	\end{minipage}
        }\hspace{-5mm}
        \subfigure[Python]{
	\begin{minipage}{0.33\linewidth}
		\centering
		\includegraphics[width=0.95\linewidth]{figure/python.pdf}
		\label{python} 
	\end{minipage}
        }
\caption{Exhibition of domain divergence in PHP, Java, and Python languages.}
\label{fig:languages}
\vspace{-5mm}
\end{figure*}

Cross-language code clone detection~\cite{vislavski2018licca, nafi2019clcdsa, tao2022c4} has been attracting increasing attention over the years.
Early works detect cross-language clones by analyzing syntactic and lexical features of the source codes~\cite{vislavski2018licca,nafi2019clcdsa,cheng2017clcminer}. For example, LICCA~\cite{vislavski2018licca} parses multilingual programs into unified syntax trees through the SSQSA platform~\cite{rakic2015extendable}, and measures similarities of their syntax trees; CLCD-SA~\cite{nafi2019clcdsa} analyzes 9 types of syntactic features across different languages to detect clones; CLCMiner~\cite{cheng2017clcminer} 
mines clones from revision histories of different languages. 
Although training-free, most of these approaches have application limitations due to manually customized rules and service conditions.
Besides, their performances are usually mediocre since they heavily depend on lexical overlap or exterior syntactic features.
With the rise of deep neural networks, pre-trained language models (PLMs) trained on the multilingual corpus have 
achieved impressive performances on various downstream tasks~\cite{tao2022c4, zhang2022learning, feng2020codebert} due to their strong capability in programming representation and understanding. Researchers turn to neural networks to detect cross-language clones. Some works~\cite{guo2022unixcoder} directly apply PLMs to detect cross-language clones without fine-tuning. Despite mapping multilingual representations into the same space, these works ignore aligning the semantically similar representations of different languages, thus achieving suboptimal performances. To further align multilingual representations, some approaches~\cite{tao2022c4} utilize a large amount of cross-language parallel clone data to fine-tune pre-trained models. 
However, in the software system, there are insufficiently labeled cross-language clones, especially for low-resource languages, and the costs of data collection, cleaning, and modeling parallel corpora are all non-neglectable. Therefore, how to design an effectively unsupervised cross-language code clone model without
any parallel multilingual clone pairs is a meaningful problem.
Compared with supervised monolingual or multilingual clone detection, zero-shot cross-language clone detection is more challenging. The main difficulties are as follows.
First,  
\emph{the characteristics of various programming languages are discrepant.} Each language exists numerous language-specific  lexical contexts, syntax, and API terms. Figure~\ref{fig:languages} reports the domain divergence between PHP, Java, and Python languages. As shown in Figure~\ref{fig:languages}, PHP language uses ``echo'' and Python language utilizes ``print'' for outputting strings, respectively. ``Array.sort()'', ``asort()'', and ``.sort()'' are used for sorting arrays and have distinguishing usage patterns in different languages.  
If we directly utilize models to detect multilingual clones, they will achieve suboptimal performances due to recognizing these discrepancies is difficult.
Second, \emph{parallel corpus is not available for fine-tuning.} Though existing pre-trained models~\cite{tao2022c4} have achieved impressive performance on clone detection, most of them rely heavily on a large amount of parallel data to detect clones. Collecting high-quality labeled cross-language datasets is expensive and time-consuming, especially for low-resource languages. In the zero-shot cross-language clone detection scenario, there is no labeled multilingual parallel data for models to learn the semantic similarity among different languages. 
Third, \emph{how to detect code clones.} Due to the diversity of the code implementations, functionally equivalent programs usually have dissimilar implementations. In addition, some non-equivalent programs look similar in lexicon and syntax. It is difficult for models to detect functionally equivalent programs with dissimilar implementations or distinguish the non-equivalent programs that have similar implementations. 

To address the aforementioned challenges, we propose a novel approach dubbed ZC$^3$ to investigate zero-shot cross-language code clone detection. ZC$^3$ is based on the large language model CodeBERT~\cite{feng2020codebert} which is pre-trained on six programming languages, thus various languages can be represented in the same vector space. Next, we aim to align the representations among different languages in high-dimensional space without any labeled cross-language parallel corpus. We propose contrastive snippet prediction to construct an isomorphic representation space among diverse languages. Based on the isomorphic structure, we exploit domain-aware learning and cycle consistency learning to empower the model to generate representations that are aligned among different
languages, meanwhile, are diacritical among diverse functions as much as possible.

To assess the effectiveness of our ZC$^3$, we conduct extensive experiments. (1) We evaluate ZC$^3$ on four representative cross-language clone detection datasets, including the CSN$_{\rm CC}$~\cite{guo2022unixcoder}, CodeJam~\cite{nafi2019clcdsa}, AtCoder~\cite{nafi2019clcdsa}, and XLCOST~\cite{zhu2022xlcost} datasets. (2) In terms of MAP, ZC$^3$ outperforms the state-of-the-art (SOTA) baselines by up to 46.59\% absolute improvements. Concretely, compared to the SOTA methods, ZC$^3$ improve the performances from 13.37\% to 66.38\%, 24.83\% to 91.96\%, 23.85\% to 75.24\%,  and 82.09\% to 96.94\% on the four datasets, respectively. 
(3) We also evaluate ZC$^3$ to detect clones on the monolingual languages and on the unseen programming languages. ZC$^3$ achieves impressive  performance on unseen programming languages and also performs well in detecting monolingual clones.
(4) We provide the ablation results by gradually removing each part. The results demonstrate the contributions of each module. 
(5) We visualize the representations of source codes. The representational distributions are well-aligned among different languages, meanwhile, are diacritical for different types of clones. The code and data for this paper can be found at https://github.com/lairikeqiA/ZC3.

We summarize our contributions in this paper as follows:
\begin{itemize}
\item Our paper investigates zero-shot cross-language clone detection where no multilingual parallel corpus is required.
\item We propose a novel zero-shot cross-language clone detection method dubbed ZC$^3$. It designs the contrastive snippet  prediction to form an isomorphic representation structure. Based on the isomorphic space, we exploit domain-aware learning and cycle consistency learning to constrain the model to generate aligned and diacritical representations among various languages.
\item  We conduct extensive experiments on four cross-language datasets. Experimental results reveal that ZC$^3$ significantly outperforms the state-of-the-art baselines. 
\end{itemize}

%% file: 2_related.tex
\section{Background }
\subsection{Code Clone Detection}
Given a program as the query, we target to retrieve codes with the same semantics from a collection of candidates. In this paper, we do not consider the classifying scenario that classifies whether two codes are semantically equivalent or not, since the retrieval task is more practical in software systems.

According to ~\citet{roy2007survey}, similar codes can be divided into two kinds. They are similar in the program text, or semantically equivalent without being textually similar. Based on the two kinds, code clones are summarized in four types, including Type I, Type II, Type III, and Type IV. Type I: Identical source codes except for variations in white space (or layout) and comments.  Type II: Code fragments have identical syntax but can be different in identifiers, literals, types, layouts, and comments. Type III: Fragments have more discrepancy except for variations in identifiers, literals, types, layout, and comments, where statements can be removed, added, and changed. Type IV: Programs have the same semantics but can be implemented through different methods, named semantic clones. In terms of their definitions, Type I-III are relevant to textual similarity and Type IV is about functionally equivalent. 

Various traditional methods~\cite{sajnani2016sourcerercc, roy2018benchmarks, jiang2007deckard} tried to analyze and detect Type I-III. ~\citet{sajnani2016sourcerercc} proposed SourcererCC, which compares tokens and subsequences to detect clones. ~\citet{jiang2007deckard} consider only syntactic information and ignore lexical similarities of programs. 
These approaches have limited success in detecting semantic clones and considering functional behaviors. To detect semantic clones (Type IV), researchers applied deep learning approaches~\cite{zhang2019novel, yu2019neural, wei2018positive, wei2017supervised}. These methods use deep neural networks to acquire code representations, then detect clones through their representations' similarities. Among these works, however, most of them can only detect clones in the monolingual scenario. 
In this paper, the zero-shot cross-language code clone task belongs to Type IV, which is difficult to detect clones since source codes have different implementations and further obey specific syntactic rules among diverse programming languages. 

\subsection{Zero-Shot Learning}
Though having achieved satisfactory performances on various downstream tasks, deep learning methods~\cite{EditSum, CodeEditor, tao2023eveval, zhang2022learning, li2022sk2}, especially pre-trained language models, are data-hungry. 
The performance of these models is strongly dependent on a large amount of parallel corpus. Models usually achieve suboptimal performance without sufficient examples. In practice, parallel data is rare and collecting them is time-consuming. To address the above issue, researchers~\cite{wu2019zero} begin to investigate to zero-shot learning. It is a machine-learning technology that aims to adapt a model to a new task or a new domain without any labeled parallel examples. 

In the software system, some PLMs~\cite{SkCoder, Self-collaboration, CODEP, zhang2023implant, zhang2022does} utilize large-scale networks and a large amount of data to learn satisfactorily semantic representations for programs. Then, they are directly applied to detect clones based on the representational space without fine-tuning. Although performing in the zero-shot setting, most of these models usually achieve poor performance or can only detect clones in the same programming language. Compared with these models, ZC$^3$ faces more challenges since it needs to align semantically similar representations among different languages without any annotated cross-language parallel corpus.

%% file: 3_method.tex
\section{Methodology: \textbf{ZC$^3$}}~\label{model}
ZC$^3$ aims to align representations of different languages on the high-dimensional space as much as possible, then detect cross-language clones based on the aligned vector space. Our proposed 
ZC$^3$ designs the contrastive snippet prediction to construct an isomorphic representation structure for different languages. Based on the isomorphic space, domain-aware learning and cycle consistency learning are exploited to empower the model to generate
well-aligned representations among different languages that meanwhile are diacritical for different types of clones. In this section, we first briefly formulate code clone detection based on PLMs and then describe our proposed ZC$^3$ method.

\begin{figure}[!t]
\centering
  \setlength{\abovecaptionskip}{0.1cm} 
	\vspace{-0.35cm} 
	\subfigtopskip=0pt 
	\subfigbottomskip=1pt 
	\subfigcapskip=0pt 
\includegraphics[width=0.8\linewidth]{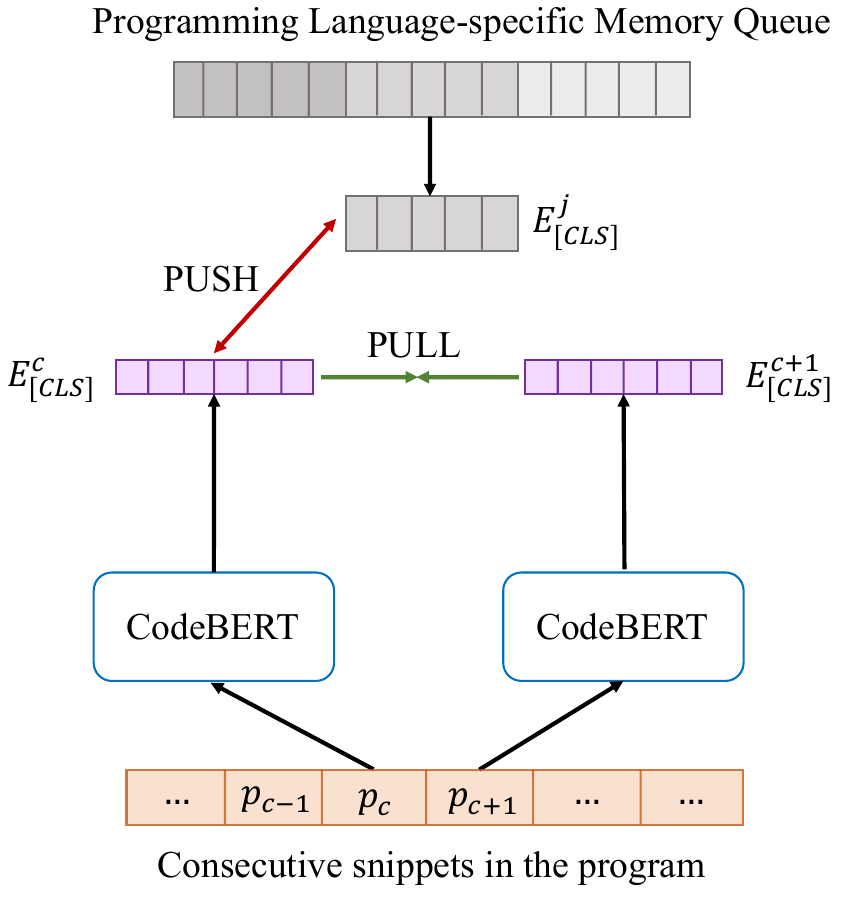}
\caption{Schematic depiction of the contrastive snippet prediction (CSP). }
\vspace{-5mm}
\label{fig:zc3-1}
\end{figure}

\subsection{Preliminary} \label{preliminary}
Given a specific program as the query, we focus on retrieving semantically equivalent codes from a candidate pool, instead of  classifying whether two programs are clones or not, since the retrieval scenario is more realistic in software systems. Thus, zero-shot cross-language code clone detection is defined as follows.
It learns a robust matching model without any cross-language parallel data and retrieves the matched codes from target languages given a source-language program. 

Recently, pre-trained programming language models, such as CodeBERT~\cite{feng2020codebert} and GraphCodeBERT~\cite{guo2020graphcodebert}, have been widely used in many code-related tasks and shown impressive results. 
In this paper, we also consider building our model based on PLMs and select CodeBERT~\cite{feng2020codebert} as the basic model since it is pre-trained on a large amount of unlabeled multilingual data and can map representations of different languages to a unified space. 
Given a source-language program as the query ${s}$, we first concatenate ${s}=\{s_1, s_2, ..., s_n\}$ with special tokens ${[CLS]}$ and ${[SEP]}$, which can be formulated as $\{{[CLS]}, s_1, s_2, ..., s_n, {[SEP]}\}$. It is tokenized into a sequence of tokens and fed into the CodeBERT. Then CodeBERT outputs a semantically embedding sequence
${S}=\{{E}^s_{{[CLS]}}, {E}^s_2,..., {E}^s_{|S|}\}$ where ${|S|}$ is the length of tokens. Meanwhile, we use the same way to process target-language fragments ${t}$ and acquire its representational sequence ${T}=\{{E}^t_{{[CLS]}}, {E}^t_2,..., {E}^t_{|T|}\}$. In this paper, we treat  ${E}^s_{{[CLS]}}$ and ${E}^t_{{[CLS]}}$ as the aggregated representations of ${s}$ and ${t}$. Finally, the semantic similarity of the two programs is formulated as: 
\begin{equation}\label{eq:cos}
\begin{aligned}
{sim}(s, t) = {cos}({E}^s_{{[CLS]}}, {E}^t_{{[CLS]}}) = \frac{{{E}^s_{{[CLS]}}}^T {E}^t_{{[CLS]}}}{ \lVert{{E}^s_{{[CLS]}}}\rVert \lVert{{E}^t_{{[CLS]}}}\rVert}
\end{aligned}
\end{equation}
where ${sim}(s, t)$ denotes the cosine similarity between the query ${s}$ and the candidate code ${t}$.

\subsection{Contrastive Snippet Prediction}\label{3.2}

Cross-language dense retrieval tasks require the model to map semantically related sequences to similar positions in representational space, but existing approaches only focus on mapping multilingual sequence pairs with the same meaning to similar embeddings~\cite{wu2022unsupervised}. To mitigate this phenomenon, we propose a new pre-training task named Contrastive Snippet Prediction (CSP).
The CSP task targets to construct isomorphic representation space among different languages by modeling the snippet relations in programs. In the isomorphic space, the code snippets with the same semantics in different languages have similar relationships to other monolingual code snippets. The CSP task is illustrated in Figure\ref{fig:zc3-1}.

\textbf{CSP Objective.} CSP task aims to establish the relations of code snippets. Formally, a functional program is a sequence of code snippets, where each snippet contains several consecutive lines. For each center code snippet, we define the snippets in the window centered on it as functional snippets. Given a center code snippet $p_c$, the CSP task needs to select the correct functional snippet $p_f$ from a randomly sampled snippet pool. 
With the CSP task, the model could assess the mutual information $I(p_f|p_c)$ of snippet relations.

Specifically, a program consists of a series of code snippets $(p_1, p_2,...,p_n)$. For each center snippet $p_c$, its functional snippet set is $F(p_c)=\{p_f|c-w <= f <= c+w, f \neq c\}$. $w$ is the radius of the window that represents the maximum snippet distance between the center snippet $p_c$ and its functional snippets $p_f$. Next, CodeBERT is applied to encode $p_c$ and $p_f$, and acquires $E^c_{[CLS]}$ and $E^f_{[CLS]}$ as their representations. We then model the relations of center snippet $p_c$ and other snippets with the contrastive loss. 
Following SimCLR~\cite{chen2020simple}, the learning objective $\mathcal{L}_{CSP}$ of the CSP task is formulated as:

\begin{equation}
\begin{aligned}
\mathcal{L}_{CSP} = & - \bigg[log \frac{e^{s(E^c_{[CLS]}, E^f_{[CLS]})/\tau}}{\sum_{E^n_{[CLS]} \in \mathbb{N}_{CSP}}e^{s(E^c_{[CLS]}, E^n_{[CLS]})/\tau}} \bigg]
\end{aligned}
\end{equation}
where $\tau$ is a temperature parameter. $s(\cdot)$ calculates the cosine similarity of two vectors. $\mathbb{N}_{CSP}$ represents the randomly selected snippets out of the functional snippet set $F(p_c)$. 


Based on the CSP task, our model constructs the isomorphic representation among different languages by learning the snippet relations. The learned representational space for different languages will have good cross-language properties after it became isomorphic by the task.

\textbf{Programming Language-specific Queue.} 
As demonstrated by~\cite{oord2018representation}, the bound of contrastive loss descends as the number of negative samples increases. We design a language-specific queue to expand the negative sample number under the limited batch size.  
In particular, the language-specific queue only stores the embedding of snippets within the same programming language. In the training process, given a language-specific query, negative samples are randomly selected only from the candidate queues that have the same language as the query. The queue is maintained in FIFO (First-In-First-Out) manner. The newest mini-batch embedding is put into the queue, meanwhile, the oldest mini-batch representations are taken out. With each iteration, language representations and parameters of the model are optimized.

\begin{figure}[!t]
\centering
\includegraphics[width=\linewidth]{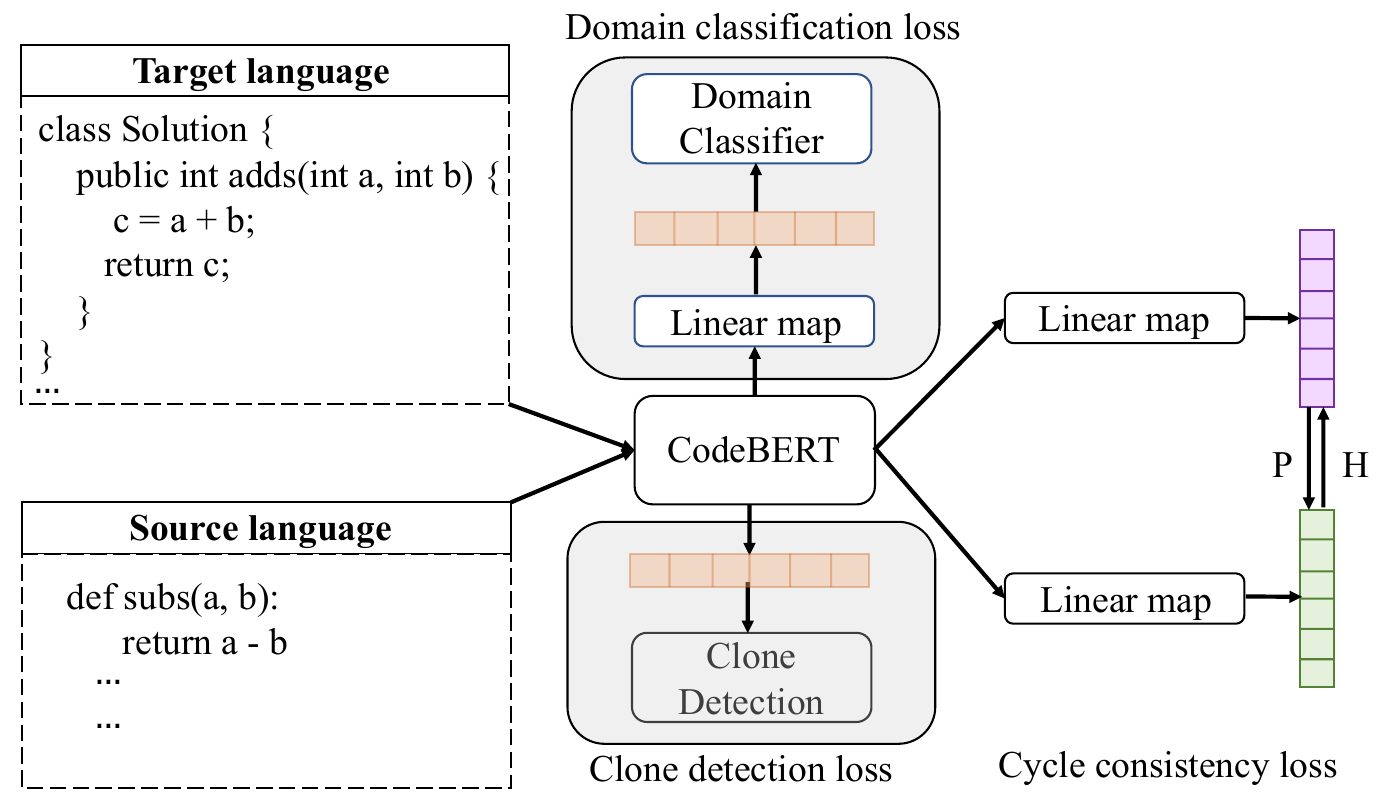}
\caption{The overview of adversarial clone detection.}
\vspace{-5mm}
\label{fig:adv-clone}
\end{figure}

\subsection{Adversarial Clone Detection}\label{3.3}
After acquiring isomorphic distributions, in order to further constrain the model to generate aligned representations, we explore domain-aware learning and cycle consistency learning at the function level. 
Based on this, we introduce a clone detection task to teach the model for detecting  clones. Finally, the representational distributions of ZC$^3$ are well-aligned among different languages, meanwhile, are diacritical for different types of clones. Figure~\ref{fig:adv-clone} illustrates the 
overview of the adversarial clone detection.

\subsubsection{Domain-Aware Learning}

To further align multilingual representations, we introduce the domain-aware learning module.
In the procedure, we add a domain classifier layer on the encoder to predict the language labels of programs, and the parameters of the encoder are optimized to maximize the domain classifier's loss by adversarial training. 
Formally, we feed a source-language program or a target-language code into CodeBERT and obtain its aggregated representation $E_{[CLS]}$. Then $E_{[CLS]}$ is fed into a gradient reversal layer (GRL) ~\cite{ganin2016domain} to reduce the domain discrepancy of $E_{[CLS]}$ for the source domain and target domain. During the forward propagation, GRL acts as an identity function but during the backpropagation, GRL multiplies the incoming gradient by a negative factor $-\lambda$ which reverses the gradient direction. We can formulate GRL as a ``pseudo-function" $G_{\lambda}(x)$ by the following two equations:
\begin{equation}
\begin{aligned}
G_{\lambda}(x)= x
\end{aligned}
\end{equation}
\begin{equation}
\begin{aligned}
\frac{\partial G_{\lambda}(x)}{\partial x} = -\lambda I
\end{aligned}
\end{equation}
where $\lambda = \frac{2}{1+exp(-\mu \frac{t}{T})}$ is a constant. $T$ is the maximum optimizing step, $t$ is the current training step, and $\mu$ is a hyper-parameter. 

Then we feed GRL's outputs into the domain classifier layer which is a non-linear transformation layer parameterized as: 
\begin{equation}
\begin{aligned}
f_{dc}(x) = \mathrm{softmax}(W_{dc}G_{\lambda}(E_{[CLS]})+b_{dc})
\end{aligned}
\end{equation}

The target is to minimize the cross-entropy for all data from both the source language and the target language:
\begin{equation}
\begin{aligned}
\mathcal{L}_{dc} = - \frac{1}{N_s+N_t} \sum_{i=1}^{N_s+N_t}y^i_{dc}logf_{dc}(x^i) \\
+ (1-y^i_{dc})log(1-f_{dc}(x^i))
\end{aligned}
\end{equation}
where $y_{dc}^i \in \left\{0,1 \right\}$ is the ground truth language label. $N_s$ and $N_t$ are the numbers of training examples in both languages.

Learning with GRL is adversarial such that the domain classifier is optimized to increase the ability to distinguish different languages, while the encoder learns representations to reduce the domain classification accuracy due to the reversal of the gradient. Through domain-aware learning, the multilingual representations are indistinguishable and further aligned.

\begin{figure}[!t]
\centering
\includegraphics[width=\linewidth]{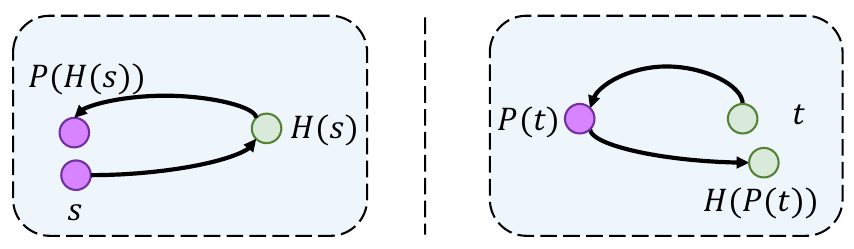}
\caption{ The demonstration of cycle consistency learning that constrains that if the model translates a program from one language to the other and back again the program arrives where it starts: $\textbf{s} \rightarrow \textbf{H(s)} \rightarrow \textbf{P(H(s))} \approx \textbf{s} $, and $\textbf{t} \rightarrow \textbf{P(t)} \rightarrow \textbf{H(P(t))} \approx \textbf{t} $.}
\vspace{-4mm}
\label{fig:number}
\end{figure}

\subsubsection{Cycle Consistency Learning}
CSP task targets learning snippet relations and constructing the isomorphic structure in the embedding space. At the function-individual level, however, the model is not constrained to generate representations that are both aligned and diacritical. Therefore, cycle consistency learning \cite{zhu2017unpaired} is designed to restrict the model to produce aligned and discriminative embeddings at the function level among different languages. The cycle consistency contains two parts, namely forward cycle consistency and back cycle consistency. For a source-language sample, the cycle can be able to bring it back to the original semantic function in the source language as much as possible, which is called forward cycle consistency. Similarly, we utilize the backward cycle consistency to map a target-language code back into its original meaning.
The forward cycle consistency can be formulated as:
\vspace{-1mm}
\begin{equation}
\begin{aligned}
F_s: s\rightarrow H(s) \rightarrow P(H(s)) \approx s
\end{aligned}
\end{equation}
\vspace{-1mm}
The backward cycle consistency is represented as:
\begin{equation}
\begin{aligned}
B_t: t\rightarrow P(t) \rightarrow H(P(t)) \approx t
\end{aligned}
\end{equation}
\vspace{-5mm}

The $H(\cdot)$ maps representations from the source language to the target language, and another linear operation $P(\cdot)$ maps representations in the reverse direction. $H(\cdot)$ and $P(\cdot)$ are both trainable linear transformations, which have the same structure but share no parameters. In the case of the $P(\cdot)$ module,  it can be formulated as following:
\begin{equation}
\begin{aligned}
P(x) = w_{2}(w_{1}x+b_{1})+b_{2}
\end{aligned}
\end{equation}
where $w_k$ and $b_k$ are training parameters.

We incentivize this behavior using a cycle consistency loss:
\begin{equation}
\begin{aligned}
\mathcal{L}_{cyc} = \mathbb{E}_{s\in S} [|P(H(s)), s |] + \mathbb{E}_{t\in T} [|H(P(t)), t |]
\end{aligned}
\end{equation}
where $|\cdot|$ is the L1 loss. $S$ and $T$ are the samples of the source language and target language, respectively.

\subsubsection{Code Clone Detection}
Based on the aligned vector space, we use the labeled monolingual data to learn effective clone ability. The task requires the model to identify semantically equivalent programs by measuring the matching degree between the code query and the candidate code. Precisely, we feed the code query and a candidate snippet into the encoder, respectively, and obtain their aggregated representations $E^q_{[CLS]}$ and $E^r_{[CLS]}$. Then their matching degrees can be calculated by cosine similarity as formulated in Formula~\ref{eq:cos}. Finally, the cross-entropy loss function is utilized as the training objective for the task:
\begin{equation}
\begin{aligned}
\mathcal{L}_{cl} = - \frac{1}{N_{m}} \sum_{i=1}^{N_{m}}y^i log(sim(E^{q,i}_{[CLS]}, E^{r,i}_{[CLS]}))
\end{aligned}
\end{equation}
where $N_{m}$ is the number of labeled monolingual examples.

In total, the adversarial code clone procedure is optimized by minimizing the following loss:
\begin{equation}
\begin{aligned}
\mathcal{L} = \mathcal{L}_{cl} + \alpha \mathcal{L}_{dc} +\beta \mathcal{L}_{cyc}
\end{aligned}
\end{equation}
where $\alpha$ denotes the weight for domain classifier loss and $\beta$ is the weight for the cycle consistency loss.

%% file: 4_experiments.tex
\section{Study Design}

To evaluate the effectiveness of our approach, we conduct a large-scale study to answer three research questions. In this section, we describe the details of our study, including datasets, metrics, and baselines.

\subsection{Research Questions}
Our study aims to answer three research questions (RQ). In RQ1, we compare our ZC$^3$ to advanced code clone baselines on four representative cross-language datasets. In RQ2, we evaluate how ZC$^3$ performs in the unseen programming languages. In RQ3, we verify the performance of ZC$^3$ on the monolingual code clone detection scenario. 

\textbf{RQ1: How does} ZC$^3$ \textbf{perform compared to the SOTA unsupervised cross-language code clone baselines?} We evaluate ZC$^3$ on four representative datasets. Considering that our approach aims to detect clones in the unsupervised setting, we compare our ZC$^3$ with the advanced unsupervised cross-language clone detection baselines.

\textbf{RQ2: What is the performance of} ZC$^3$ \textbf{in the unseen programming languages?} 
In this paper, 
we also assess whether ZC$^3$ can effectively detect cross-language clones in the unseen languages.

\textbf{RQ3: How does} ZC$^3$ \textbf{perform on the monolingual code clone detection?} 
Our approach is designed for zero-shot cross-language clone detection and aligns representational distributions for different languages. In this RQ, we evaluate whether ZC$^3$ performs well on the monolingual clone detection.

\begin{table}[!t] 
\renewcommand{\arraystretch}{1.1}
\centering
\caption{Statistics of XLCoST for contrastive snippet prediction. \#num means the number of snippets and programs. \#lines and \#tokens denote the average lines and tokens. } 
\resizebox{0.95\linewidth}{!}{
\begin{tabular}{lcccccc}
\toprule
\multirow{2}{*}{} & \multicolumn{3}{c}{Snippet-level}   & \multicolumn{3}{c}{Program-level}  \\ 
\cmidrule(lr){2-4} \cmidrule(lr){5-7}  
& Python & Java & Avg & Python & Java & Avg \\ \midrule  
\#num  & 81207 & 91089 & 86148 &9263 &9623 & 9443 \\
\#lines  & 2.41 &3.71 &3.06  &20.54 &34.93 & 27.74  \\
\#tokens  & 21.63 & 24.1 & 22.87 &188.5 & 227.1 & 207.8  \\  \bottomrule
\end{tabular}
}
\vspace{-5mm}
\label{tab:xlcost}
\end{table}

\subsection{Datasets and Metrics} \label{datasets}
To construct isomorphic embedding space, we use the training set of XLCoST dataset~\cite{zhu2022xlcost} for contrastive snippet prediction. XLCoST is collected from GeeksForGeeks~\footnote{https://www.geeksforgeeks.org/}, which is a website containing many programming solutions for data structures and algorithm problems. XLCoST contains 7 different programming languages including C++, Java, Python, C\#, Javascript, PHP, and C. 
Each program is divided into several consecutive code snippets, where each snippet describes a complete semantic meaning such as if-statement and for-statement. The statistics of XLCoST are given in Table~\ref{tab:xlcost}.

We use the dataset released by ~\citet{guo2022unixcoder} for adversarial code clone in Section \ref{3.3}. We name it CSN$_{\rm CC}$ in this paper.
CSN$_{\rm CC}$ is collected from the CodeSearchNet corpus~\cite{puri2021project} and contains 11,744, 15,594, and 23,530 function-level programs in Ruby, Python, and Java languages, respectively. Each program can solve one of 4,053 problems. We utilize the first 3,500 problems for adversarial code clone and randomly select other 500 problems to evaluate ZC$^3$. 

We evaluate our model on four cross-language code clone datasets including the Google CodeJam dataset~\cite{nafi2019clcdsa}, AtCoder dataset~\cite{nafi2019clcdsa}, XLCoST corpus~\cite{zhu2022xlcost}, and CSN$_{\rm CC}$ dataset~\cite{guo2022unixcoder}. CodeJam comes from Google’s programming competition\footnote{http://code.google.com/codejam}. There are 265 problems and each problem has several solutions with Python or Java languages. These problems are divided into train, valid, test sets in the ratio of 8:1:1. We only use the test set to evaluate ZC$^3$. The second dataset is the AtCoder corpus, which is collected from the AtCoder programming website\footnote{https://atcoder.jp/} in Japan. There are 1,115 classes of problems in Python and Java. The test set has 115 problems with 1,365 programs. The third dataset is the XLCoST dataset~\cite{zhu2022xlcost}. Different from the above two datasets, there is only one solution for each problem in peer language. It contains 11,028, and 10,622 programs for Java and Python. 
We use the test set of CSN$_{\rm CC}$~\cite{guo2022unixcoder} as the fourth dataset. There are 500 problems and all of them come from the public GitHub repositories~\footnote{https://github.com/}.

To verify the monolingual clone ability, we further introduce two monolingual clone detection datasets, i.e., the BigCloneBench dataset~\cite{wang2020detecting} and the POJ-104 dataset~\cite{mou2016convolutional}. 
BigCloneBench was first proposed by \citet{svajlenko2014towards} and we use the filtered version provided by \citet{wang2020detecting}. It has 9,134 codes in total and 912 programs in the test set. We group the test programs according to their semantic functions and remove categories that only contain one program. Finally, 623 programs are remained and divided into 8 functional classes.  
POJ-104 is collected from a pedagogical programming open judge (OJ) system that automatically judges the correctness of the submitted source code for specific problems. It contains 104 problems and each problem has 500 student-written C programs. Following \citet{lu2021codexglue}, we divide these problems into 64, 16, and 24 classes for training, validation, and testing. 

Following previous works~\cite{guo2022unixcoder, lu2021codexglue}, we use MAP as the metric. MAP is the mean of average precision scores, which is evaluated for retrieving similar samples given a query.



\begin{table}[t!]
  \centering
  \caption{Statistics of six code clone detection datasets on the different split sets.}
    \begin{tabular}{lcccccc}
    \toprule
    \multicolumn{1}{l}{\multirow{2}[4]{*}{Multilingual}} & \multicolumn{3}{c}{Java} & \multicolumn{3}{c}{Python} \\
\cmidrule{2-7}          & \multicolumn{1}{c}{Train} & \multicolumn{1}{c}{Valid} & \multicolumn{1}{c}{Test} & \multicolumn{1}{c}{Train} & \multicolumn{1}{c}{Valid} & \multicolumn{1}{c}{Test} \\
    \midrule
        CodeJam   &  3461  &  390 & 433 &2846  & 258  &372  \\
       AtCoder   &  11824  &   1642  & 1408 & 11631  & 1337   & 1365 \\
       CSN$_{\rm CC}$  & 15478 & 4952 &  3100  & 8081 & 3229 & 3197 \\
        XLCoST  & 9623  &  472 &  887  &  9263   &  472 & 887 \\
    \midrule
    \multicolumn{1}{l}{\multirow{2}[4]{*}{Monolingual}} & \multicolumn{3}{c}{C} & \multicolumn{3}{c}{Java} \\
\cmidrule{2-7}          & \multicolumn{1}{c}{Train} & \multicolumn{1}{c}{Valid} & \multicolumn{1}{c}{Test} & \multicolumn{1}{c}{Train} & \multicolumn{1}{c}{Valid} & \multicolumn{1}{c}{Test} \\ 
\midrule
BigCloneBench  &  -- &--  &-- &7,310 & 912  &623   \\
POJ-104  &32000  &8000 & 12000 & --  &  -- &  --  \\
    \bottomrule
    \end{tabular}%
    \vspace{-5mm}
  \label{tab:dataset-2}%
\end{table}%

\begin{table*}[!t] 
\renewcommand{\arraystretch}{1.1}
\centering
\caption{Evaluation results of zero-shot cross-language clone detection on the Python $\rightarrow$ Java setting and the Java $\rightarrow$ Python setting of the AtCoder, CodeJam, and XLCoST datasets.} 
\resizebox{0.9\linewidth}{!}{
\begin{tabular}{lcccccc}
\toprule
\multirow{2}{*}{} & \multicolumn{2}{c}{AtCoder}   & \multicolumn{2}{c}{CodeJam}  & \multicolumn{2}{c}{XLCoST} \\ 
\cmidrule(lr){2-3} \cmidrule(lr){4-5} \cmidrule(lr){6-7} 
& Python$\rightarrow$Java & Java$\rightarrow$Python & Python$\rightarrow$Java & Java$\rightarrow$Python & Python$\rightarrow$Java & Java$\rightarrow$Python  \\ \midrule
CodeBERT  & 3.81 &  2.98 & 8.33 & 8.84 & 3.36  & 1.98 \\
GraphCodeBERT & 10.23 & 6.05 & 14.95 & 15.10  & 35.32  & 23.19\\
UniXcoder  & 28.81  & 20.86 & 24.42 & 23.28 & 84.25  & 79.93\\ \midrule
CodeBERT$_{\rm MCC}$ &  89.36  &  84.72 &   63.45 & 68.66 & 92.87  & 92.63 \\
CodeBERT$_{\rm CSP}$ &  89.84   & 85.56  & 67.17 &  69.83  & 93.62  & 94.05 \\
CodeBERT$_{\rm DAL}$ &  91.89 &  87.28 & 69.30 &  72.74 & 95.36  &  96.11 \\ \midrule
ZC$^3$    & \textbf{92.25  ($\uparrow$ 3.23\%)}   &  \textbf{91.67 ($\uparrow$ 8.21\%)} &  \textbf{73.92 ($\uparrow$ 16.50\%)}  & \textbf{76.57 ($\uparrow$ 11.52\%)} & \textbf{96.96 ($\uparrow$ 4.40\%)} & \textbf{96.92 ($\uparrow$ 4.63\%)}  \\  \bottomrule
\end{tabular}
}
\vspace{-2mm}
\label{tab:result1}
\end{table*}

\begin{table*}[!t] 
\renewcommand{\arraystretch}{1.1}
\centering
\caption{Evaluation results of zero-shot cross-language code clone detection on the X $\rightarrow$ Y settings of CSN$_{\rm CC}$ dataset. The X and Y represent Ruby, Java, or Python languages.}
\resizebox{0.85\linewidth}{!}{
\begin{tabular}{lccccccc}
\toprule
\multirow{2}{*}{} & \multicolumn{2}{c}{Java}   & \multicolumn{2}{c}{Python}  & \multicolumn{2}{c}{Ruby}  & \multicolumn{1}{c}{Avg}  \\ 
\cmidrule(lr){2-3} \cmidrule(lr){4-5} \cmidrule(lr){6-7}  \cmidrule(lr){8-8}    
& Python & Ruby & Java & Ruby & Python & Java &  \\ \midrule
CodeBERT  &  0.83  &  0.96 & 1.23 & 1.30  & 1.47  & 1.03 & 1.14\\
GraphCodeBERT   & 2.01 & 2.04  & 4.37 & 4.78 & 5.03 & 4.54  &3.79\\
UniXcoder & 8.49 & 7.83 &  13.41 &  18.26  & 20.34  & 11.90 &13.37  \\ \midrule
CodeBERT$_{\rm MCC}$ & 54.26  &50.97  & 58.13  & 67.04  & 65.52  &  55.78 &58.62  \\
CodeBERT$_{\rm CSP}$ & 55.70 & 52.42  & 59.46  & 67.81 & 67.36  &  56.47 &59.87\\
CodeBERT$_{\rm DAL}$ &  57.92 & 55.12 & 61.54 & 69.60 & 67.85 &58.22 &61.71 \\ \midrule
ZC$^3$    & \textbf{63.69}   &  \textbf{63.28} &  \textbf{64.05}  & \textbf{72.18} & \textbf{72.32} & \textbf{62.76}  &\textbf{66.38} \\ 
 &($\uparrow$ \textbf{17.38\%}) & ($\uparrow$ \textbf{24.15\%}) & ($\uparrow$ \textbf{10.18\%}) & ($\uparrow$ \textbf{7.67\%})  & ($\uparrow$ \textbf{10.38\%})  & ($\uparrow$ \textbf{12.51\%})  & ($\uparrow$ \textbf{13.24\%}) \\
\bottomrule
\end{tabular}
}
\vspace{-5mm}
\label{tab:result2}
\end{table*}

\subsection{Compared Models} \label{baselines}
Considering that this paper investigates zero-shot cross-language clone detection, we compare ZC$^3$ with the unsupervised cross-language clone detection baselines as follows:
\begin{itemize}
\item \textbf{CodeBERT}~\cite{feng2020codebert}: The model is an encoder-only structure, which is pre-trained in six programming languages with various advanced pre-training tasks.
\item \textbf{GraphCodeBERT}~\cite{guo2020graphcodebert}: The model considers the relation of various variables and proposes a graph-guided masked attention to incorporate structure information of codes.
\item \textbf{UniXcoder}~\cite{guo2022unixcoder}: It is a unified pre-trained model that uses mask attention matrices to control behaviors. Besides, the model utilizes multi-modal contrastive learning and cross-modal generation tasks to align representations among different languages and provides SOTA results.

\item  \textbf{CodeBERT}$_{\textbf{MCC}}$: Fine-tuning CodeBERT on the monolingual code clone pairs in Python and Java languages.
\item  \textbf{CodeBERT}$_{\textbf{CSP}}$: Post-training CodeBERT with the contrastive snippet prediction task as described in Section~\ref{3.2}. Then it is continually fine-tuned on the monolingual clone pairs in the same way with CodeBERT$_{\rm{MCC}}$.
\item  \textbf{CodeBERT}$_{\textbf{DAL}}$: The model executes the adversarial domain-aware learning based on the CodeBERT$_{\rm CSP}$. The model can be regarded as an ablation model where the cycle consistency learning is removed from our ZC$^3$.
\end{itemize}

As described in Section~\ref{compared baselines}, we notice some zero-shot cross-language methods~\cite{vislavski2018licca, cheng2017clcminer} utilize manually customized rules to detect cross-language clones at Type I-III. We think these studies have a different research focus from this paper (Type IV). Besides, previous works~\cite{tao2022c4} have demonstrated that PLMs outperform them by a large margin. In this paper, we do not directly compare ZC$^3$ to these studies and leave them for future work.

\subsection{Implementation Details}
We use CodeBERT-base~\cite{feng2020codebert} to initialize our model. It is an encoder-only backbone having a 12-layer Transformer with 12 attention heads, 64 head sizes, and 768 hidden sizes. For contrastive snippet prediction, we set the window size to 5, in other words, the largest distance between the snippet center and the functional snippet is 2. The size of the language-specific queue is 128. During adversarial clone detection, $\mu$ is set as 0.01. The number of source-language samples and target-language samples is a ratio of 1:2. $\alpha$ and $\beta$ are set to 1, respectively. The batch size is 32. The maximum length of the source and target programs are both 512. The optimizer is an AdamW optimizer with an initial learning rate 2e-5.

\section{Results and Analyses}

\begin{table*}[!t] 
\renewcommand{\arraystretch}{1.1}
\centering
\caption{The performance of different models on unseen C++, C\#, JavaScript, PHP, and C programming languages. CodeBERT$_{\rm MCC}$ and ZC$^3$ are trained on the Java and Python languages.} 
\resizebox{0.9\linewidth}{!}{
\begin{tabular}{lcccccccccc}
\toprule
\multirow{2}{*}{} & \multicolumn{5}{c}{Python}   & \multicolumn{5}{c}{Java}  \\ 
\cmidrule(lr){2-6} \cmidrule(lr){7-11}  
& C++ & C\# & JavaScript & PHP & C & C++ & C\# & JavaScript & PHP & C \\ \midrule
CodeBERT  & 2.95 & 3.48 & 2.73 &4.25 & 1.79 &3.67  & 3.09 & 2.44 &5.24 &15.93  \\
CodeBERT$_{\rm MCC}$ & 91.37 & 93.18 & 88.92 &92.36 & 94.49 & 93.78  &97.12  & 84.02 &94.92 &96.59  \\
ZC$^3$  & \textbf{94.61}  & \textbf{96.65} &\textbf{95.35}  & \textbf{95.92} &\textbf{95.88} & \textbf{95.81} & \textbf{98.82} & \textbf{95.67} & \textbf{97.70}  & \textbf{98.83} \\ 
 & ($\uparrow$ \textbf{3.55\%})  & ($\uparrow$ \textbf{3.72\%}) &($\uparrow$ \textbf{6.74\%})  & ($\uparrow$ \textbf{3.85\%}) & ($\uparrow$ \textbf{1.47\%}) & ($\uparrow$ \textbf{2.16\%})  &  ($\uparrow$ \textbf{1.75\%}) & ($\uparrow$ \textbf{12.68\%}) &  ($\uparrow$ \textbf{3.98\%})  & ($\uparrow$ \textbf{2.32\%}) \\ \bottomrule
\end{tabular}
}
\vspace{-4mm}
\label{tab:unseen result}
\end{table*}

In our first research question, we evaluate the performance of our ZC$^3$ with respect to advanced code clone approaches.

\textbf{RQ1: How does} ZC$^3$ \textbf{perform compared to  the SOTA
unsupervised cross-language code clone baselines?} 

\textbf{Setup.} 
We evaluate baselines (Section \ref{baselines}) and our ZC$^3$ on four cross-language datasets (Section \ref{datasets}). The evaluation metric is described in Section \ref{datasets}, i.e., the MAP score. For the metric, higher scores represent better performance.

\textbf{Results.} 
Table~\ref{tab:result1} reports the performances of all methods on the AtCoder, CodeJam, and XLCoST datasets. Table~\ref{tab:result2} shows the results on CSN$_{\rm CC}$. The percentages in parentheses are the relative improvements compared to CodeBERT$_{\rm MCC}$. 

\textbf{Analyses.} 
(1) \ul{ZC$^3$ achieves the best performance among all baselines with significant improvements.} Concretely, compared with the SOTA baseline, UniXcoder, ZC$^3$ improves the MAP score from 24.83\% to 91.96\%, 23.85\% to 75.24\%, and 82.09\% to 96.94\% on the three datasets, respectively. 
(2) \ul{Our approach effectively aligns representational distributions among different languages.} Despite significant improvements of training on the monolingual labeled corpus, the gap between CodeBERT$_{\rm MCC}$ and ZC$^3$ is still up to 6.1\% on average of the three datasets. That demonstrates our designed contrastive snippet learning, domain-adversarial learning, and cycle consistency learning accelerate to construct a well-aligned vector space. 
(3) \ul{Constructing an isomorphic representation distribution is beneficial to multilingual clone detection.} By introducing contrastive snippet learning, CodeBERT$_{\rm CSP}$ further improves CodeBERT$_{\rm MCC}$ on the MAP score. The CSP task effectively grasps the snippet relations in different languages.
(4) \ul{Domain-aware learning and cycle consistency learning improves performance significantly.} 
Compared to CodeBERT$_{\rm CSP}$, ZC$^3$ acquires 4.70\% absolute improvements on MAP and achieves the best performance. That verifies that constraining the model to generate aligned and diacritical representations at the function level is essential and effective for cross-language clone detection.

To illustrate the generalization ability of our ZC$^3$, we further conduct experiments on CSN$_{\rm CC}$ dataset that is collected from GitHub. From Table~\ref{tab:result2}, we can find that ZC$^3$ gets 53.01\% average absolute improvements on the six settings in terms of MAP, and the trends are generally consistent with the results in Table~\ref{tab:result1}, which further demonstrates the effectiveness and generalization ability of our method. In addition, although the CSP task is only trained in Java and Python languages, the procedure can also improve the performance on the Java $\rightarrow$ Ruby setting and the Python $\rightarrow$ Ruby setting. 

\begin{tcolorbox}[enhanced,colback=gray!5!white,colframe=gray!75!black,drop fuzzy shadow southwest]
  \textbf{Answer to RQ1: ZC$^3$ achieves the best results among all baselines. In particular, ZC$^3$ acquires 67.13\%, 51.30\%, 14.85\%, and 53.01\% absolute improvements on four datasets at MAP. The significant improvements prove our approach can produce well-aligned representational distributions and is a promising approach for zero-shot cross-language code clone detection.} 
\end{tcolorbox}

\begin{table}[!t] 
\renewcommand{\arraystretch}{1.1}
\centering
\caption{Evaluation results of the monolingual code clone detection on the POJ-104 dataset and the BigCloneBench dataset.}
\resizebox{0.75\linewidth}{!}{
\begin{tabular}{lcc}
\toprule
\multirow{2}{*}{} & \multicolumn{1}{c}{POJ-104 } &\multicolumn{1}{c}{BigCloneBench}   \\ 
\midrule
CodeBERT$_{\rm FT}$  &  86.42  &  65.37  \\
ZC$^3$  &  86.27  &  62.59   \\
\bottomrule
\end{tabular}
}
\vspace{-6mm}
\label{tab:result3}
\end{table}

\textbf{RQ2: What is the performance of} ZC$^3$ \textbf{in the unseen programming languages?} 

\textbf{Setup.} We evaluate how ZC$^3$ performs in the unseen languages. Precisely, we use the model trained on Python and Java as described in Section~\ref{model} to evaluate the Python $\rightarrow$ X and Java $\rightarrow$ X settings, where X $\in \left\{\textrm{C++, C\#, JavaScript, PHP, C}  \right\}$ is never seen during the training procedure.

\textbf{Results.} The results are shown in Table~\ref{tab:unseen result}.  The percentages in parentheses mean the relative improvements from CodeBERT$_{\rm MCC}$ to ZC$^3$. 

\textbf{Analyses.}
(1) \ul{Our ZC$^3$ can work well in the unseen programming languages.} As shown from results, ZC$^3$ achieves 95.68\% on the Python $\rightarrow$ X setting and 97.37\% on the Java $\rightarrow$ X setting in terms of MAP. Besides, ZC$^3$ outperforms CodeBERT$_{\rm MCC}$ by a large margin in all settings. It brings 3.87\% and 4.58\% relative improvements on the two settings compared with CodeBERT$_{\rm MCC}$.
That demonstrates that the aligned representation space constructed by contrastive snippet prediction and adversarial clone detection is universal and can be translated into other unseen languages.
(2) \ul{The improvements on the Java $\rightarrow$ X setting is higher than the Python $\rightarrow$ X setting.} The reason might be that Java, C++, C\#, JavaScript, and PHP are object-oriented programming languages, and their syntax and code morphology are similar. 

\begin{tcolorbox}[enhanced,colback=gray!5!white,colframe=gray!75!black,drop fuzzy shadow southwest]
  \textbf{Answer to RQ2: Our ZC$^3$ achieves impressive performances in unseen programming languages. That demonstrates that our approach works well and can be applied to other unseen languages.} 
\end{tcolorbox}

\begin{figure*}[htbp]
	\centering
        \setlength{\abovecaptionskip}{-5.mm}
        \subfigure{
	\begin{minipage}{0.25\linewidth}
		\centering
		\includegraphics[width=1.1\linewidth]{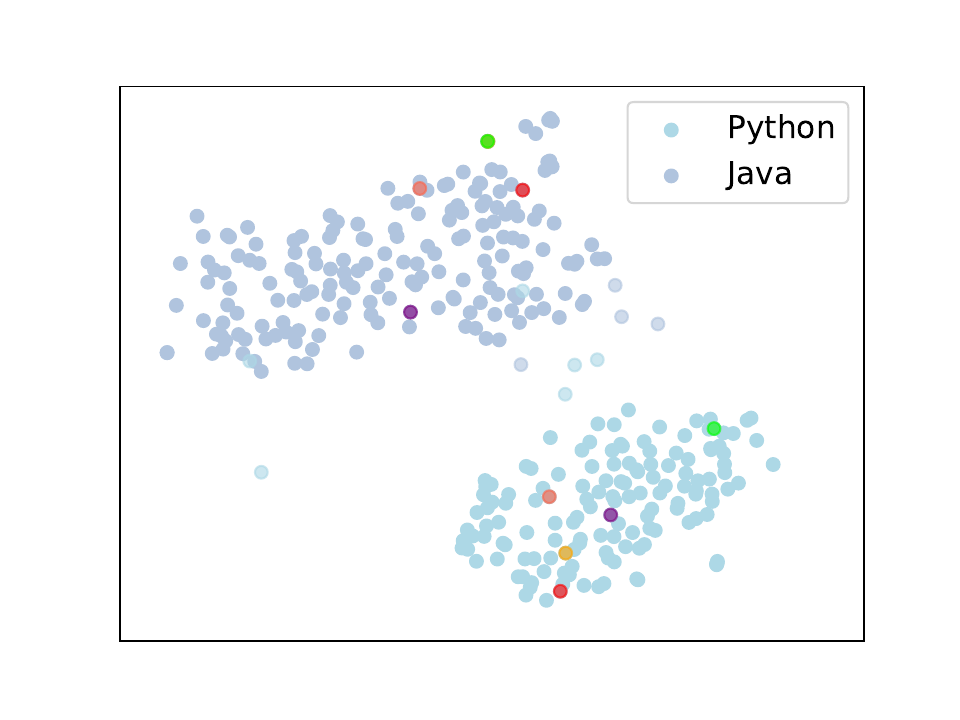}
		\label{CodeBERT}
	\end{minipage}
        }
        \hspace{0.85cm}
        \subfigure{
	\begin{minipage}{0.25\linewidth}
		\centering
		\includegraphics[width=1.1\linewidth]{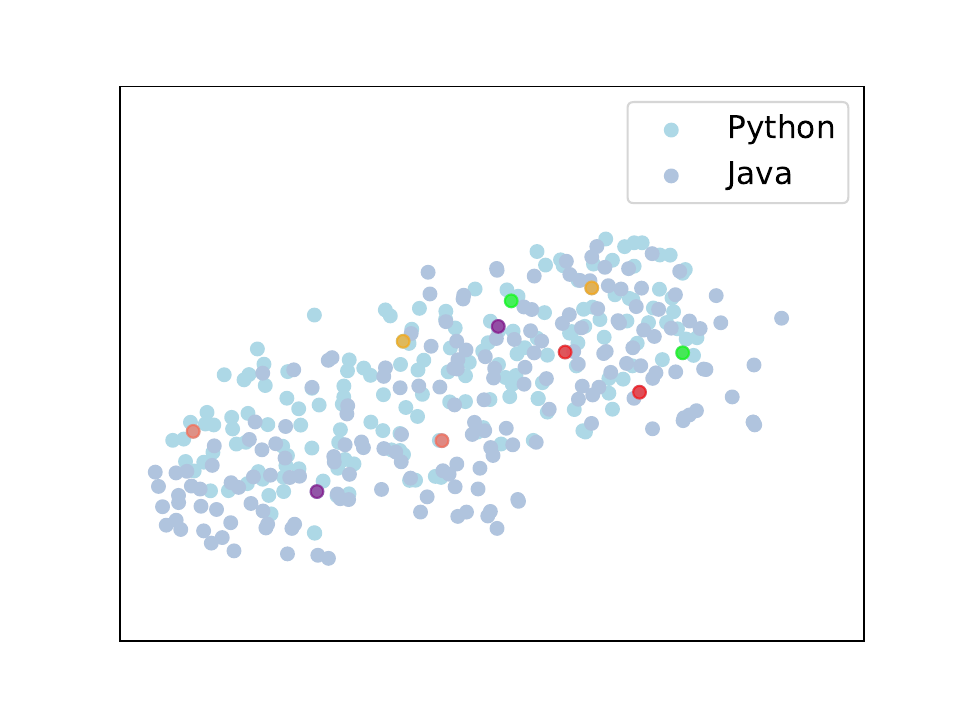}
		\label{DAL}
	\end{minipage}
        }
        \hspace{0.85cm}
       \subfigure{
	\begin{minipage}{0.25\linewidth}
		\centering
		\includegraphics[width=1.1\linewidth]{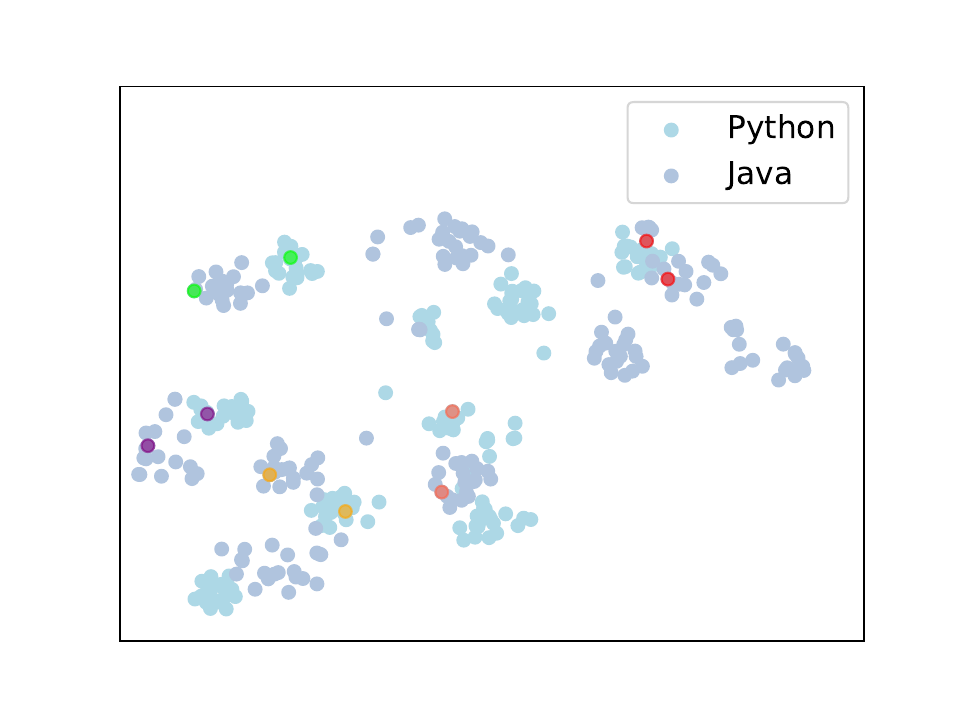}
		\label{ZC3} 
	\end{minipage}
        }
        \vspace{2mm}
\caption{Comparision of t-SNE visualization in CodeBERT (left), CodeBERT$_{\rm DAL}$ (middle), and ZC$^3$ (right). Each data point in the figure is an example. Data points with the same color represent a clone pair. Different colors mean different types of clones.}
\vspace{-5mm}
\label{fig:features}
\end{figure*}

\textbf{RQ3: How does} ZC$^3$ \textbf{perform on the monolingual code clone detection?} 

\textbf{Setup.} We assess the performance of our ZC$^3$ on the monolingual clone detection scenario. For a fair comparison, we train CodeBERT on the training set of POJ-104 and BigCloneBench, respectively, and name it CodeBERT$_{\rm FT}$.

\textbf{Results.} We present experimental results of CodeBERT$_{\rm FT}$ and ZC$^3$ in Table~\ref{tab:result3}.

\textbf{Analyses.} (1) \ul {Our ZC$^3$ performs well on the monolingual code clone detection.} In particular, ZC$^3$ achieves 86.27\% and 62.59\% on MAP at the POJ-104 dataset and the BigCloneBench dataset, respectively. This means that our approach not only achieves SOTA results on zero-shot cross-language clone detection but also performs well in detecting monolingual clones. (2) \ul{The performances of ZC$^3$ are competitive to CodeBERT$_{\rm FT}$.} CodeBERT$_{\rm FT}$ is trained on the training set of the two datasets and then detects clones on the test set, which is a supervised procedure and can be treated as the upper bound. Compared to CodeBERT$_{\rm FT}$, the results of our ZC$^3$ are only slightly lower, which demonstrates ZC$^3$ is effective enough in the monolingual clone detection scenario. In addition, considering that ZC$^3$ is also trained on the monolingual clone corpus, we also explore why ZC$^3$ is lower than CodeBERT$_{\rm FT}$. The reason might be that our labeled example size is smaller than the counterpart of CodeBERT$_{\rm FT}$ and the representational alignment losses affect the model to focus on learning Java clone detection ability.

\begin{tcolorbox}[enhanced,colback=gray!5!white,colframe=gray!75!black,drop fuzzy shadow southwest]
  \textbf{Answer to RQ3: ZC$^3$ performs well on the monolingual clone detection scenario, and its performance is close to the counterpart of the supervised method, e.g., CodeBERT$_{\rm FT}$. In particular, ZC$^3$ acquires 86.27\% and 62.59\% at MAP on both datasets. } 
\end{tcolorbox}

\begin{figure}[!t]
\centering
\includegraphics[width=0.8\linewidth]{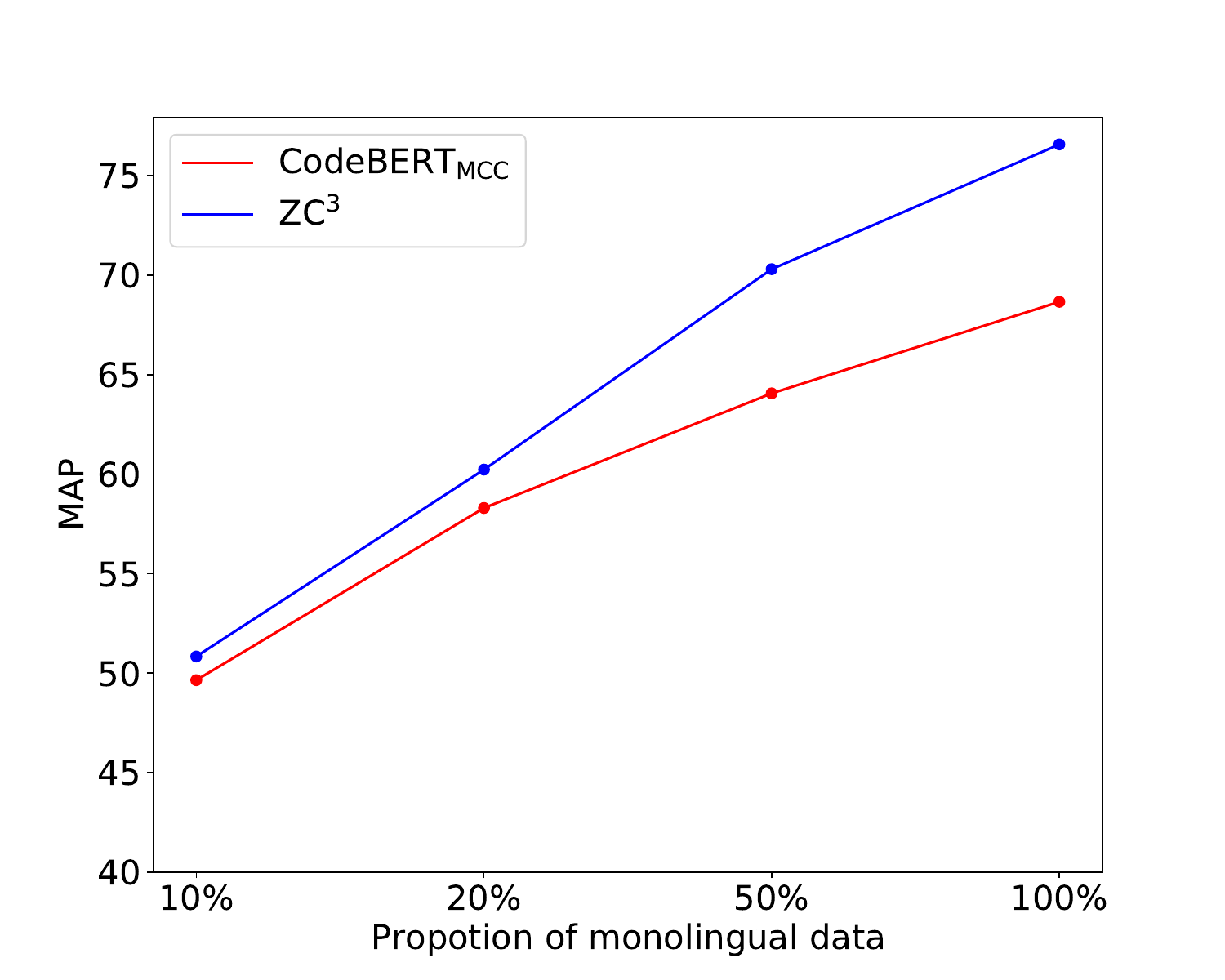}
\vspace{-2mm}
\caption{Effects of the number of monolingual labeled clone pairs about CodeBERT$_{\rm MCC}$ and ZC$^3$ on the Python $\rightarrow$ Java setting of CodeJam dataset.}
\vspace{-5mm}
\label{fig:number}
\end{figure}

\section{Discussion}

\subsection{Impacts of the Number of Monolingual Clone Pairs}
We further study the impacts of the number of labeled clone samples in the monolingual language. Figure~\ref{fig:number} reports how the performances of ZC$^3$ and CodeBERT$_{\rm MCC}$ change with respect to different proportions of labeled monolingual clone samples on the CodeJam dataset. We can find that the performances of both models increase monotonically with the increase of monolingual training samples. In addition, ZC$^3$ always outperforms CodeBERT$_{\rm MCC}$ in all cases, and the improvements become more significant with the increase of labeled training samples.

\subsection{Visualization of Features}
We further perform the visualization of example representation vectors on CodeBERT, CodeBERT$_{\rm DAL}$, and ZC$^3$.  We randomly select ten classes of Python-Java clone pairs from the test set of the CodeJam dataset. Then we feed them into models and obtain their representations. The dimension of these representations are reduced from 768 to 2 by t-SNE~\cite{van2008visualizing}, and each sample is mapped into a point at Figure~\ref{fig:features}.
From these samples, we randomly highlight five clone pairs on different classes, and the two highlighted cycles in the same color belong to a clone pair. Different colors mean different types of clones. 
As shown in the left subfigure, the distribution of two languages has a clear border, which demonstrates that CodeBERT is sensitive to the language type. By introducing the CSP task and domain-aware learning, the features of Python and Java languages are well entangled but the same type of clones are not aligned as shown in the middle subfigure. After cycle consistency learning, both languages not only entangle together but also each type of code clone is well clustered, which verifies that the model is constrained to generate aligned and diacritical features as much as possible.

\subsection{Effects of Window Size and the Queue}
The window size and the queue are important elements in the contrastive snippet prediction procedure. We investigate how the two elements affect the performance of our model. As shown in Table~\ref{tab:size}, with the increasing of window size and the queue size, ZC$^3$ acquires consistent improvements. Between the two factors, queue size has a greater influence on the performance of our model.


\subsection{Threats to Validity}

There are two main threats to the validity of our work.

\textbf{The generalizability of our experimental results.} To mitigate this threat, we carefully design our experimental datasets, baselines, and metrics. For the datasets, we follow previous studies ~\cite{nafi2019clcdsa, zhu2022xlcost, guo2022unixcoder} and use four representative cross-language clone detection datasets. The four datasets cover different programming languages (i.e., Java, Python, and Ruby, etc.) and come from different domains. In addition, we also introduce two popular monolingual code clone corpora, POJ-104 and BigCloneBench, to assess our approach's ability on monolingual clone detection. To verify the superiority of ZC$^3$, we select a series of advanced zero-shot cross-language clone detection models as our baselines for the comparison. They are the representative PLMs for understanding programs in the past three years. 
For the metric, following existing works~\cite{guo2022unixcoder, lu2021codexglue}, we select a widely used MAP metric to evaluate all methods. It is defined as the mean of average precision scores and is applied to evaluate for retrieving similar samples given a query. Besides, we execute each method three times and report the average experimental results.

\textbf{The implementation of models.}
It is widely known that deep neural models are sensitive to the implementation details, including network architectures and hyper-parameters. In this paper, we need to execute all baselines and our approach. For the baselines, we apply the source code and parameters published by their original papers ~\cite{feng2020codebert, guo2020graphcodebert, guo2022unixcoder}. Then, we ensure the models' performances are comparable with their reported results. For our approach, we use the mainstream neural network (e.g., CodeBERT) as our backbone as described in Section \ref{preliminary}. To select suitable hyper-parameters, we implement a small-range grid search on several hyper-parameters (i.e., snippet window size $w$, adversarial clone detection parameter $\mu$, etc), and leave other hyper-parameters including batch size and learning rate the same with baselines. Thus, there might be room to tune network architectures and more hyper-parameters of our approach for more improvements.

%% file: 7_conclusion.tex
\section{Related Work}

\begin{table}[t]
  \centering
  \caption{Effects of window size and the queue.}
    \begin{tabular}{ccc}
    \toprule
     & \multicolumn{1}{c}{Python$\rightarrow$Java} & \multicolumn{1}{c}{Java$\rightarrow$Python} \\
    \midrule
    Window Size &       &  \\
    2     &  73.19 &76.15\\
    3     &   73.37 & 76.24  \\
    5     & 73.92 & 76.57\\
    \midrule
    Queue &       &  \\
    32    & 70.81 & 75.40 \\
    64    & 71.54 & 75.86 \\
    128   & 73.92 & 76.57 \\
    \bottomrule
    \end{tabular}%
    \vspace{-6mm}
  \label{tab:size}%
\end{table}%

\subsection{Code Clone Detection}

Existing clone detection approaches can mainly be divided into traditional methods and learning-based methods. Traditional approaches aim to solve Type I-III clone detection by considering token and syntactic information. 
Token-aware works treat the code into a sequence of tokens and match tokens between the query and candidate code. CCFinder~\cite{kamiya2002ccfinder} and SourcerCC~\cite{sajnani2016sourcerercc} convert codes into a regular form and identify clones through a token-by-token matching algorithm. They only consider textual information and rely heavily on language-related rules and lexical vocab, which limits its practical application. 
Syntactic clone methods use Abstract Syntax Trees (AST) to compute the similarity of codes. 
Deckard~\cite{jiang2007deckard} maps the subtrees into numerical vectors in Euclidean space and clusters them. 
Clones are detected by comparing the whole tree. Similar to CCFinder, Deckard also needs to define specific rules for different languages.


Learning-based approaches mainly focus on Type IV clones, which detect semantically equivalent programs. The early deep neural-based clone detection method is CCLearner~\cite{li2017cclearner}. CCLearner trains a classifier and then utilizes it to detect clones from a candidate pool. ~\citet{wei2017supervised} propose a neural framework that exploits the lexical and syntactical information for acquiring representations and computes semantical similarities between programs. \citet{zhang2019novel} design an AST-based model, named ASTNN, that splits the whole AST into a number of small subtrees. Then RNN encodes these subtrees and produces a vector for detecting clones.
Recently, various PLMs are proposed and applied to detect clones. CodeBERT~\cite{feng2020codebert} is an early-designed PLMs that is pre-trained on 6 programming languages and computes the similarity of learned code representations for clone detection. GraphCodeBERT~\cite{guo2020graphcodebert} designs a graph-guided masked attention to encode the syntactic structure of codes and achieves impressive on the task. Inspired by contrastive learning~\cite{chen2020simple}, a series of approaches~\cite{ding2021contrastive, jain2020contrastive} utilize it for learning function-level code representations. These methods augment a given sample by constructing a similar counterpart, then force models to recognize similar programs from a candidate pool. However, these methods mainly focus on detecting monolingual clones.


\subsection{Cross-lingual Representation Alignment} \label{compared baselines}

Cross-lingual representation alignment aims to map semantically equivalent but linguistically different contents into similar positions in a unified vector space. Recently, multilingual representation has been studied in many natural language tasks~\cite{devlin2018bert, li2020unsupervised, lample2019cross, jiang2020cross}. 
XLM~\cite{lample2019cross} and XLM-R performs cross-lingual pre-training by introducing the translation language task on multilingual datasets. 


When it comes to cross-language alignment for programs, existing models are mainly divided into two kinds. One group of models uses labeled cross-lingual clones for supervised learning. 
CLCDSA~\cite{nafi2019clcdsa} proposes a parallel dataset that contains labeled cross-lingual clones. Then it uses the dataset to train a neural network for detecting clones. C4~\cite{tao2022c4} uses a large amount of parallel multilingual data for contrastive learning, which is a competitive method for supervised cross-language clone detection. 
Another school of research is unsupervised cross-language clone detection models that contain rule-based and learning-based methods.
The rule-based methods contain  LICCA~\cite{vislavski2018licca} and CLCMiner~\cite{cheng2017clcminer}. LICCA~\cite{vislavski2018licca} depends on SSQSA's high-level representation and meanwhile requires programs to be the same length. CLCMiner~\cite{cheng2017clcminer} needs the clones from revision histories. These requirements limit their 
practical application. The learning-based models refer to PLMs, such as CodeBERT~\cite{feng2020codebert}, GraphCodeBERT~\cite{guo2020graphcodebert}, and UniXcoder~\cite{guo2022unixcoder}. They are pre-trained in different languages and directly applied to detect multilingual clones.
Despite these methods can perform zero-shot clone detection on multilingual languages, their performances are sub-optimal since they either focus on lexical similarity or ignore modeling the semantical relations among different languages.

\section{Conclusion}
During software development, there is an increasing demand for zero-shot cross-language clone detection. 
This paper proposes a novel model ZC$^3$. We design a contrastive snippet prediction task to learn snippet relations and construct an isomorphic structure for different languages. Based on the isomorphic space, we exploit domain-aware learning and cycle consistency learning to further constrain the model to generate aligned representations among various languages that meanwhile are diacritical for different types of clones. Experimental results on four cross-language clone datasets demonstrate that our proposed approach achieves excellent performance on zero-shot cross-language clone detection.



\section*{Acknowledgment}
This research is supported by the National Natural Science Foundation of China under Grant Nos. 62192731, 62192733, 62192730, 61751210, 62072007, and 61832009. We also would like to thank all the anonymous reviewers for constructive comments and suggestions to this paper.
